\documentclass[twocolumn,showpacs,preprintnumbers,amsmath,amssymb,aps]{revtex4}

\usepackage{graphicx}
\usepackage{dcolumn}
\usepackage{bm}

\begin{document}

\preprint{}

\title{Unveiling the dynamics of the universe}

\author{L. Arturo Ure\~na--L\'opez}%
 \email{lurena@fisica.ugto.mx}
\affiliation{%
Instituto de F\'isica de la Universidad de Guanajuato, C.P. 37150,
 Le\'on, Guanajuato, M\'exico.}%

\date{\today}

\begin{abstract}
We present two methods for describing, from a pedagogical point of
view, the solutions of Einstein's equations applied to a homogeneous
and isotropic universe. In the first method, we define an effective
gravitational potential of the universe, and the second method makes
use of the fact that we can define a dynamical system of
equations. The methods are applied to different cosmological models
whose properties are discussed in turn. The ultimate intention is to
provide simple examples to be revised in a first Cosmology course for
undergraduates.
\end{abstract}

\maketitle

\section{Introduction}
In any introductory course on Cosmology, some questions undoubtedly
arise once the students become familiar with the equations describing
the evolution of a homogeneous and isotropic universe. These are: a)
which is the universe made of?; b) what is the curvature of the
universe? c) how did the universe expand in the past according to the
answers in a) and b)?.

It is interesting to note that present technology has allowed the
humankind to partially answer those questions with some important
accuracy\cite{Lambda,Conley:2006qb,wiki3}. However, such an
improvement in our knowledge of the universe will take still some
years to reach the textbooks for undergraduates. 

In this paper, we briefly review the basic equations of the
evolution of the universe, and describe how the latter evolves
according to its material contents and curvature. It is not our
aim to give an exhaustive study of the gravitational dynamics that
arises from the equations of Einstein's General Relativity (GR), but
rather to give some tools to help students to understand the richness
of the solutions of the Einstein's equations in the case of a
homogeneous and isotropic universe. In other words, we will focus our
attention in the answer to question c) above.

In the same line, our intention is also to provide simple pedagogical
exercises, and to solve them with the help of analytical methods which
are of widespread use in the specialized literature. The chosen
examples in this paper are appropriate, in our opinion, for a first
undergraduate course on Cosmology.

We shall not discuss how we can determine the material content of
the universe and its true nature through observations, nor how its
curvature is measured. We will only mention briefly how such
quantities are obtained, and give the reader some interesting references and
internet links where more detailed information can be found. However,
non-experts will find a comprehensible summary of modern Cosmology
in\cite{wiki3}, and some interesting questions reviewed and answered
in\cite{wright,Lopez-Corredoira:2003uc}.

The present manuscript is organized as follows. In
Sec.~\ref{sec:mathback}, we review the metric quantities that describe
a homogeneous and isotropic spacetime, and how they are influenced by
the material content of the universe through the equations of
Einstein's General Relativity. We describe the types of matter we
shall consider, and how each one is classified according to their
equation of state.

We shall introduce the so-called density parameters, which
measure the relative contribution of each component to the total
material content of the universe. These parameters will play a central
role in the subsequent calculations. Due to its particular importance,
we will take some space to discuss the significance of the curvature's
density parameter.

In Sec.~\ref{sec:gravpot}, we describe how the Friedmann equation can
be used to define an effective gravitational potential of the
universe, so that we can visualize the expansion of the universe in a
similar fashion as one describes the motion of a single particle in
Classical Mechanics.

In Sec.~\ref{sec:dynasys}, we use again the Friedmann equation and
the equations of motion of each single fluid to obtain a set of
differential equations that allows us to see the expansion of the
universe as the solutions of a dynamical autonomous system. The latter
formalism will be used to see whether there is any attractor behavior
in the cosmological solutions.

The formalisms developed in the previous two sections is applied to
three particular examples in Sec.~\ref{sec:expansion}. These are: the
actual standard cosmological model, also known as the Concordance
Model (CM); Einstein's static model of the universe; and a CM with an
arbitrary component of what we shall call dark energy.

Finally, Sec.~\ref{sec:conclusions} is devoted to conclusions and
general comments.

\section{Mathematical background}
\label{sec:mathback}
We start with the so-called \emph{Cosmological Principle (CP)}, first
proposed by Einstein, which states that the universe we live in is
homogeneous and isotropic in large scales. By large scales we mean
scales much larger than the size of a typical galaxy.

The CP is a working hypothesis, and it is the simplest assumption we
can make about the properties of the spacetime of the universe as a
whole. The interested reader can find a more detailed discussion on
the historical and philosophical relevance of Einstein's CP
in\cite{peebles93,Ellis:2006fy}.

\subsection{Space geometry}
The remarkable thing is that the CP suffices to fix the metric of the
spacetime a homogeneous and isotropic universe must have. It is called
the Friedmann-Lemaitre-Robertson-Walker (FLRW) metric, a metric with
constant curvature, whose line element is usually written as (in units
with $c=1$)
\begin{equation}
ds^2 = g_{\mu \nu} dx^\mu dx^\nu = -dt^2 +a^2(t) \left[
  \frac{dr^2}{1-kr^2} +r^2 d\Omega^2 \right] \, , \label{flrw}
\end{equation}
where the coordinate $t$ is called the cosmic time, $a(t)$ is the
(time-dependent) scale factor, and $k$ is the curvature constant. 

The spatial part of the metric is written in terms of
\textit{comoving} coordinates, as they remain fixed as the universe
expands. Actually, the \textit{physical} and \textit{comoving}
distances at a given time $t$ are related through $x_\textrm{ph}=a(t)
x_\textrm{com}$.

Notice that $d\Omega$ is the usual spherical solid angle element, and
$r$ is a comoving radial coordinate. In general, for a fixed $r$ and
$t$, we see that the circumference of a circle is equal to $C=2\pi r$,
and the surface of a sphere is $S=4\pi r^2$\cite{landau}. But,
the proper radius of these objects, for fixed $\theta$ and $\varphi$,
is given by
\begin{equation}
  \ell = \int^r_0 \frac{dr}{\sqrt{1-kr^2}} = \left\{ 
\begin{array}{cc}
\frac{1}{\sqrt{k}} \sin^{-1}(\sqrt{k}r) & \textrm{for} \, k > 0 \, , \\
r & \textrm{for} \, k = 0 \, , \\
\frac{1}{\sqrt{-k}} \sinh^{-1}(\sqrt{-k}r) & \textrm{for} \, k < 0 \, . 
\end{array}
\right.
\end{equation}
As a consequence, we see that for for $k=0$, we recover the usual
Euclidean relations $C/\ell = 2\pi$ and $S/\ell^2 = 4\pi$. However, if
the curvature constant is positive $k>0$, then $C/\ell < 2\pi$ and
$S/\ell^2 < 4\pi$. The opposite happens for a negative curvature $k <
0$, and hence $C/\ell > 2\pi$ and $S/\ell^2 > 4\pi$.

The expansion of the universe is encoded in the time derivative
 (which we denote by a dot) of the scale factor, with which one
 defines the \textit{Hubble parameter} $H\equiv \dot{a}/a$, named
 after the astronomer Edwin Hubble. Its importance comes from the fact
 that nearby galaxies recede with a velocity proportional to their
 (physical) distance from us, as $v = H
 x_\textrm{ph}$\cite{Hubble:1929ig,Davis:2003ad,Padmanabhan:2006kz}.

\subsection{Cosmological equations}
The equations of motion of the universe are given by Einstein's GR,
which is our current fundamental theory of
gravitation\cite{landau,peebles93,rindler01,padh94,Ellis:2006fy,padh94,liddle90}.
For this, we need to specify the kind of matter the universe is made of,
and give its corresponding energy-momentum tensor. The usual choice is
that of a homogeneous and isotropic perfect fluid, which is well
described and defined only by its energy density $\rho(t)$ and its
isotropic pressure $p(t)$.

There are two equations coming from GR, relating the geometry and material
contents of the universe, and there is another equation coming from
the conservation of the energy-momentum
tensor\cite{peebles93,rindler01,Padmanabhan:2006kz,padh94,liddle90}.
For a heuristic approximations to the cosmological equations
see\cite{Jordan:2003tt,McCauley:2000}. The equations of motion are
\begin{subequations}
\label{einstein}
\begin{eqnarray}
\dot{H} &=& - 4\pi \, G \sum^n_i \left( \rho_i + p_i \right) +
\frac{k}{a^2} \, ,
\label{einsteina} \\
\dot{\rho}_i &=& -3H \left( \rho_i + p_i \right) \, , \label{einsteinb}
\end{eqnarray}
\end{subequations}
together with the so called Friedmann (constraint) equation
\begin{equation}
H^2 = \frac{8\pi \, G}{3} \sum^n_i \rho_i -\frac{k}{a^2} \, . \label{fried}
\end{equation}

We have allowed for the existence in the universe of more than one
perfect fluid; with the explicit restriction that they do not interact with
each other except gravitationally. This means that each of the perfect
fluids obeys a separate conservation equation, namely
Eq.~(\ref{einsteinb}).

\subsection{Types of matter}
Only $(n+1)$ of the above $(n+2)$ equations are independent, and we
need $(n)$ extra equations in order to solve for the $(2n+1)$ unknowns
$a(t)$, $\rho_i(t)$ and $p_i(t)$. The needed equation is what is
called an \emph{equation of state} relating the energy density and
pressure of each perfect fluid. we will follow the common wisdom and
assume a \emph{barotropic} equation of state in the form
\begin{equation}
p_i = (\gamma_i -1) \rho_i \, ,
\end{equation}
where $\gamma_i$ is the equation of state of the $i$-th fluid. In this
case, Eq.~(\ref{einsteinb}) integrates to
\begin{equation}
\rho_i = \rho_{i,0} a^{-3\gamma_i} \, , \label{eq:1}
\end{equation}
where $\rho_{i,0}$ is an integration constant. The most usual values
of $\gamma_i$ are: 
\begin{itemize}
\item Radiation and relativistic particles, $\gamma_r =4/3$ and
  $\rho_r \sim a^{-4}$;

\item Pressureless matter (dust or non-relativistic particles),
  $\gamma_m =1$ and $\rho_m \sim a^{-3}$; 

\item Cosmological constant ($\Lambda$ or vacuum energy),
  $\gamma_\Lambda =0$ and $\rho_\Lambda \sim
  \textrm{const}$\cite{Alam:2004jy} 
\end{itemize}y
However, it is widely accepted that the allowed range can be $2\geq
\gamma \geq 0$\footnote{Nevertheless, the case for $\gamma < 0$ cannot
be discarded. A fluid with such an equation of state is called
\emph{phantom energy}, see for
instance\cite{Caldwell:1999ew,Scherrer:2004eq}.}.

Eqs.~(\ref{einstein}) and~(\ref{fried}) shows that the curvature of
the universe contributes to the dynamics. Actually, we can
think of curvature as a kind of special perfect fluid with
$\gamma_k=2/3$, so that $\rho_k \sim a^{-2}$, as it is indeed the
case.

\subsection{Critical energy density and density parameters}
The Friedmann equation~(\ref{fried}) can be rewritten in the
dimensionless form
\begin{equation}
1=\sum^n_i \frac{\rho_i}{\rho_c} -\frac{\rho_k}{\rho_c} = \sum^n_i
\Omega_i + \Omega_k \, . \label{fried1}
\end{equation}
We have defined in here the \emph{critical energy density} $\rho_c$ and the
\emph{density parameters} $\Omega_i$ as follows
\begin{equation}
\rho_c \equiv \frac{3 H^2}{8 \pi \, G} \, , \quad \Omega_i \equiv
\frac{\rho_i}{\rho_c} \, , \quad \Omega_k \equiv \frac{-k}{a^2 H^2}
\, . \label{densyparams}
\end{equation}
The density parameter for the curvature component $\Omega_k$ is
defined together with the negative sign, so that it is on equal
footing with respect to the density parameters of the material
components\cite{Liddle:1998ew,Sanders:2004xi}.

It can be seen from Eq.~(\ref{fried1}) that if the whole material
content of the universe equals the critical energy density, $\sum
\rho_i = \rho_c$, then the curvature of the universe should be zero;
that is, it should correspond to a flat universe. Likewise, if $\sum_i
\rho_i > \rho_c$ ($\sum_i \rho_i < \rho_c$) then the universe has a
closed (open) spatial geometry.

The critical energy density depends on the value of the Hubble
parameter, and thus is a time-dependent quantity. However, the current
value of $\rho_{c,0}$ can be measured directly from the redshift of
nearby galaxies and other objects, see\cite{Conley:2006qb,Reid:2002kp}
for recent results.

On the other hand, each of the density parameters $\Omega_i$ shows the
\emph{relative} contribution of each type of matter to the critical energy
density at any time. The allowed range for any density parameter is
then $0 \leq \Omega_i \leq 1$, and $\Omega_i =1$ means that the
evolution of the universe is \emph{dominated} by the $i$-th
fluid. Whenever this happens we usually speak of the $i$-th fluid
dominated era in the evolution of the universe.

Also, $\Omega_{i,0}$ would represent the current relative contribution
of $\rho_{i,0}$ to the current critical energy density
$\rho_{c,0}$. Hereafter, a subscript $'0'$ will denote current values
for any quantity.

\subsection{The curvature of the universe}
There is a normalization issue about the curvature of the universe we
should be careful with. It is often said that the curvature constant
can be normalized to take three different values: a) $k=-1$, for an
\textit{open} (Hyperbolic) universe; b) $k=0$, for a \textit{flat}
(Euclidean) universe; and c) $k=1$ for a \textit{closed} (Spherical)
universe\footnote{Though the labeling may resemble that of $2$-dimensional
surfaces, we should keep in mind that the spatial part of
metric~(\ref{flrw}) refers to $3$-dimensional hypersurfaces.}. For
further simplicity, one usually finds the suggestion of normalizing
the scale factor too, so that its actual value is $a_0 =1$.

On the other hand, the curvature contributes to the energy density of
the universe, and the product $kr^2$ in metric~(\ref{flrw}) should be
dimensionless. Thus, we have two options.
\begin{itemize}
\item \emph{Dimensionless} $k$ and $r$, and a scale factor $a(t)$ with
  dimensions of length. In this case, we can normalize the curvature
  constant as mentioned above, but we should notice then that the
  scale factor cannot be normalized arbitrarily as its actual value is
  set by the curvature density parameter as $a_0 =
  |\Omega_{k,0}|^{1/2} H_0$.

\item \emph{Dimensionless} scale factor $a(t)$, $r$ with dimensions of
  length, and $k$ with dimensions of $\textrm{length}^{-2}$. Thus, the
  curvature constant is given by $k = -a^2_0 H^2_0
  \Omega_{k,0}$. Without loss of generality, we can choose to
  normalize the scale factor as $a_0=1$.
\end{itemize}
For convenience, the second option will be used throughout this paper.

The constant of curvature is given, at any time, by $k = -a^2 \Omega_k
H^2$. This is an interesting relation, since $d_H \equiv H^{-1}$,
which is called the \emph{Hubble length}, provides us of an estimate
of the size of the observable universe at any
time\cite{peebles93,Ellis:2006fy,landau,rindler01,liddle90,padh94}.
Hence, $k$ tells us of the deviation of the spatial part of
metric~(\ref{flrw}) from the flat case. For scales $r$ for which
$\sqrt{|k|} r= a \sqrt{|\Omega_{k}|} H \, r \ll 1$, the universe can
be considered to have an Euclidean space geometry.

The current value of the Hubble parameter is $H_0 = 70 \, \textrm{km}
\, \textrm{s}^{-1} \,
\textrm{Mpc}^{-1}$\cite{Conley:2006qb,Lambda,Padmanabhan:2006kz,wiki2}\footnote{$1
  \, \textrm{pc} = 3.2 \, \textrm{light-years}$.}, which implies that
the current Hubble distance (an estimation of the \textit{current}
size of our observable universe) is $d_{0,H} \simeq 4,300 \,
\textrm{Mpc}$. Also, the current value of curvature's density
parameter is $\Omega_{0,k}= 0.01$\cite{Lambda,wiki2}, which
seems to indicate we live in an (slightly) open universe.

Thus, the universe we actually see can be plainly taken as Euclidean,
since the scales at which the (current) curvature of the universe
could be appreciable in the metric at the present time ($>
d_{H,0}/\sqrt{|\Omega_{k,0}|} \simeq 43,000 \, \textrm{Mpc}$) are
larger than the distance to the farthest object we can
observe!\footnote{The current \emph{particle horizon}, which is the
  distance light has traveled since the Big Bang up to date, is of the
  order of $3.3 \, d_{H,0}$. Notice that this distance is larger than
  the age of the universe multiplied by the velocity of light, see for
  instance the discussion on this topic in\cite{wright,Davis:2003ad}.}

\section{The effective gravitational potential of the universe}
\label{sec:gravpot}
We will now present the first method to describe the expansion of the
universe according to the type of matter is made of. First, we notice
that each energy density can be given in terms of the actual value of
its corresponding density parameter as $\rho_i = \Omega_{i,0}
\rho_{c,0} \, a^{-3\gamma_i}$, see Eqs.~(\ref{eq:1})
and~(\ref{densyparams}).

Second, the Friedmann equation~(\ref{fried}) can be rewritten in the
form\cite{peebles93,padh94,rindler01,Padmanabhan:2006kz}
\begin{equation}
\frac{\dot{a}^2}{2} +V(a) = \frac{1}{2} \Omega_{k,0} \, , \label{fried2}
\end{equation}
where now a dot means derivative with respect to the dimensionless
time $\tau = H_0 t$. The cosmic time is normalized in terms of the
actual value of the so-called \emph{Hubble time} $H^{-1}_0 = 14$
Gy. The effective gravitational potential $V(a)$ explicitly reads
\begin{equation}
V(a)= - \frac{1}{2} \sum^n_i \Omega_{i,0} a^{-3\gamma_i+2} \, .
\label{gravpot}
\end{equation}

Eq.~(\ref{fried2}) resembles the conservation of energy for a particle
with ``space'' coordinate $a(t)$ and constant energy
$(1/2) \Omega_{k,0}$\footnote{The equation of
  motion~(\ref{fried2}) is well known to textbooks,
  see for example\cite{peebles93,padh94}, the detailed discussion
  in\cite{rindler01}, and the Wikipedia text in\cite{wiki1}. However, its
  pedagogical properties have not been fully exploited even though
  Eq.~(\ref{fried2}) is indeed widely used as a serious research tool
  in the specialized literature\cite{Szydlowski:2006ma}.}. We see that
a flat universe has zero total energy, and a closed (open) universe
has negative (positive) total energy. Also, we would like to stress
out that the actual values of the density parameters are not all
independent, but are related through the Friedmann constraint at the
present time,
\begin{equation}
1= \sum^n_i \Omega_{i,0} + \Omega_{k,0} \, . \label{fried3}
\end{equation}

The gravitational potential~(\ref{gravpot}) is \emph{negative}
definite, $V(a)<0$, if all of the (actual) density parameters are
\emph{positive} definite, $\Omega_{i,0} \geq 0$. This fact may imply
in some cases the presence of at least one \textit{maximum} in the
potential. This can be verified by direct calculation of the first and
second derivatives,
\begin{subequations}
\label{deriv}
\begin{eqnarray}
V^\prime (a) &=& \frac{1}{2} \sum^n_i \left( 3\gamma_i-2\right)
\Omega_{i,0} a^{-3\gamma_i+1} \, , \label{deriv1} \\
V^{\prime \prime}(a) &=& \frac{1}{a} V^\prime (a) -\frac{3}{2}
\sum^n_i (3\gamma_i -2) \gamma_i \Omega_{i,0} a^{-3\gamma_i} \,
. \label{deriv2}
\end{eqnarray}
\end{subequations}

The existence of a critical point is not a trivial thing, as it marks
the value of the scale factor at which $\ddot{a}=0$, since
Eq.~(\ref{gravpot}) is equivalent to the \emph{acceleration equation}
$\ddot{a}=-V^\prime (a)$. At $a=a_c$, the acceleration of the
universe's expansion vanishes. Therefore, the existence of a critical
point tells us that the universe should have had a decelerated and an
accelerated expansion at some stages.

When does a critical point exist? From Eq.~(\ref{deriv1}), there
cannot be a critical point if all of the equations of state $\gamma_i
> 2/3$; the universe will always decelerate in such case. In other
words, it is necessary the presence of at least one perfect fluid with
an equation of state with a value less than $2/3$ for the universe to
have an accelerated expansion.

Actually, it can be proved in the general case that if there is a
critical point $a_c$ such that $V^{\prime} (a_c)=0$, then $V^{\prime
  \prime}(a_c) < 0$. For this, let us assume that $(n-1)$ matter
fluids have an equation of state $\gamma_i > 2/3$, and that it is only
the $n$-th fluid which has $\gamma_n < 2/3$. Using the condition
$V^{\prime} (a_c)=0$ in Eq.~(\ref{deriv1}), we can write
\begin{equation}
  V^{\prime \prime}(a_c) = -\frac{3}{2} \sum^{n-1}_i (3\gamma_i -2)
  (\gamma_i - \gamma_n) \Omega_{i,0} a^{-3\gamma_i}_c \, .
\end{equation}
All the terms inside the sum are positive by assumption; therefore,
  $V^{\prime \prime}(a_c) < 0$. The final result is the same if more
  than one fluid have an equation of state $\gamma < 2/3$. From this
  we conclude that any critical point corresponds to a maximum.

\section{Dynamical cosmological system}
\label{sec:dynasys}
In the presentation of the second method, we find convenient to take
the density parameters as the dynamical variables themselves. It
is then our purpose in this section to show how to write the Einstein
equations as a \emph{dynamical autonomous system} (for a comprehensive
reading of such systems see\cite{jorgevjose};
and\cite{Copeland:1997et,Collinucci:2004iw,Urena-Lopez:2005zd,Copeland:2006wr,Coley:1999uh,dynasystems}
for applications in Cosmology).

\subsection{The general case}
Let us assume that there are $n+1$ perfect fluids present in the
universe. This is the most general case, as we have already mentioned
that the curvature itself can be seen as a special perfect fluid. The
Friedmann constraint helps us to reduce in one the number of
independent variables, i.e., it is only necessary to consider $n$
perfect fluids\footnote{If we think of the density parameters as the
  coordinates of a $n+1$-dimensional Cartesian space, the Friedmann
  equation~(\ref{fried1}) then forces the universe to 'move' on a
  particular hyperplane only.}.

Choosing the $(n+1)$-th perfect fluid to be absorbed by means of
Eq.~(\ref{fried1}), the Einstein equations for the rest of the $n$
fluids is given in terms of their corresponding density parameters as
\begin{equation}
\Omega^\prime_j = 3 \Omega_j \left[ \sum^n_{i=1} \left( \gamma_i -
 \gamma_{n+1} \right) \Omega_i -  \left( \gamma_j - \gamma_{n+1}
 \right) \right] \, , \label{dynasysa}
\end{equation}
where $i,j=1,2, \dots, n$, and $\gamma_{n+1}$ is the barotropic equation
of state of the (absorbed) $(n+1)$-th perfect fluid. The prime denotes
derivative with respect to the so-called \emph{number of $e$-foldings}
$N\equiv \ln a$. To arrive to Eq.~(\ref{dynasysa}), we have also made
use of Eq.~(\ref{einsteina}) in the form
\begin{equation}
1 + \frac{\dot{H}}{H^2} = \frac{\ddot{a}}{H^2 a} = - \frac{3}{2}
\sum^n_{i=1} \Omega_i \left( \gamma_i - \gamma_{n+1} \right) - \left(
\frac{3}{2} \gamma_{n+1} -1 \right) \, . \label{dynasysb}
\end{equation}

\subsubsection{Exact solution}
System~(\ref{dynasysa}) may appear redundant and unnecessary at first
sight. This is because the behavior of the density parameters as
functions of the number of e-foldings $N$ can be easily found. By
definition (see Eqs.~(\ref{densyparams})), the \emph{exact} solutions
of Eqs.~(\ref{dynasysa}) are
\begin{equation}
\Omega_j = \Omega_{j,0} \, \frac{e^{-3\gamma_j N}}{\sum^{n+1}_{i=1}
  \Omega_{i,0} e^{-3\gamma_i N}} \,
  , \label{dpsols}
\end{equation}
where $j=1,2,\ldots,n+1$. It is not difficult to verify that
Eqs.~(\ref{dynasysa}) directly follow from Eqs.~(\ref{fried1})
and~(\ref{dpsols}).

If one only wants to know the evolution of the density parameters,
then Eq.~(\ref{dpsols}) suffices to know all of the different stages
the universe has gone through during its evolution. See sections below
for some explicit examples.

\subsection{Fixed points and stability analysis}
However, the interesting thing to note is that Eq.~(\ref{dynasysa}) is
a dynamical autonomous system of the form $\mathbf{x}^\prime =
\mathbf{f}(\mathbf{x})$, where $\mathbf{x}= (\Omega_1,
\Omega_2,\ldots,\Omega_n )$. In this respect, Eqs.~(\ref{dynasysa})
can be seen as a complementary part of Eq.~(\ref{fried2}), since the
latter can tell us which attractor properties can be found in the
dynamical equations~(\ref{einstein}) and~(\ref{fried}).

The critical (fixed) points $\mathbf{x}_c$ of the dynamical
system~(\ref{dynasysa}) are found by solving the equations
$\mathbf{f}(\mathbf{x}_c)=0$. There are two obvious solutions.
\begin{itemize}
\item \emph{Trivial ($(n+1)$-th perfect fluid dominated) solution}, for which
  $\mathbf{x}_c=\mathbf{0}$, $\Omega_{n+1}=1$, and then the expansion of
  the universe is driven by the $(n+1)$-th perfect
  fluid. Eq.~(\ref{dynasysb}) points out that the universe has an
  accelerated expansion, $\ddot{a} > 0$, (decelerated expansion,
  $\ddot{a} < 0$) if $\gamma_{n+1} < 2/3$ ($\gamma_{n+1} > 2/3$).

  There are two important remarks we should be aware of at this point.
  \begin{itemize}
    \item If the $(n+1)$-th perfect fluid is the curvature, then
    $\gamma_{n+1} = \gamma_k = 2/3$, and then the universe expands at
    a constant rate, i.e., $\ddot{a}=0$. It should be noticed,
    however, that the existence of this critical point is forbidden by
    the Friedmann constraint~(\ref{fried1}) for the case of a
    \emph{closed} universe (if all of the density parameters are
    \emph{positive} definite).
    
    On the other hand, this trivial solution is permitted for an
    \emph{open} universe, and is better known as \emph{Milne's
      universe}\cite{peebles93,rindler01,Macleod:2005}.
    
    \item Milne's model should be distinguished from the so-called
    \emph{empty static model}\cite{rindler01}. Eqs.~(\ref{einstein}) tell
    us that an empty universe (no perfect fluid present, zero
    curvature) is indeed permitted, which will remain static,
    $a(t)=\textrm{const}$. 
    \end{itemize}

\item \emph{$j$-th perfect fluid dominated solution}, for which
  $\Omega_j=1$, and $\Omega_{i\neq j} = \Omega_n = 0$; that is, the critical
  points are $\mathbf{x}_{c,1}=(1,0,\ldots,0)$,
  $\mathbf{x}_{c,2}=(0,1,\ldots,0)$, etc. Eq.~(\ref{dynasysb})
  again indicates that the universe will accelerate (decelerate) its
  expansion if the dominant equation of state is such that $\gamma_j <
  2/3$ ($\gamma_j > 2/3$), which is the same conclusion we reached at
  in the previous section.
\end{itemize}

Once found, the stability of the critical points can be established by
a first order perturbation analysis\cite{dynasystems}, in which one
considers a small perturbation $\mathbf{u}$ in the form
$\mathbf{x}=\mathbf{x}_c+\mathbf{u}$. Hence, Eqs.~(\ref{dynasysa}) are
linearized in the form $\mathbf{u}^\prime= \mathcal{M} \mathbf{u}$,
where
\begin{equation}
\mathcal{M}_{jl} = \left. \frac{\partial f_j}{\partial x_l}
\right|_{\mathbf{x}_0} \, , \label{matrixpert}
\end{equation}
are the elements of the perturbation matrix $\mathcal{M}$.

If the eigenvalues of the matrix $\mathcal{M}$ have all negative
(positive) real parts, then the critical point is called \emph{stable}
(\emph{unstable}). If neither, it is then called a \emph{saddle} point.

Back to our case, we have to solve the set of
equations~(\ref{dynasysa}) for $\Omega^\prime_j =0$, and to study the
eigenvalues of the corresponding perturbation matrix. Let us suppose
that we want to investigate the stability of the fixed point
corresponding to the domination of the $i-$th perfect fluid,
$\mathbf{x}_{c,i}=(0,0,\ldots, \Omega_{i} =
1,\ldots,0)$.

After careful calculations, the elements of the perturbation matrix
are explicitly given by
\begin{equation}
\mathcal{M}_{jl} = 3 \delta_{jl} \left( \gamma_{i} -
  \gamma_j \right) + 3 \delta_{ji} \left( \gamma_l -
  \gamma_{n+1} \right) \, , \label{eq:3}
\end{equation}
where $\delta_{jl}$ is the Kronecker delta, $\delta_{jl}=1$ if $j=l$,
and $\delta_{jl}=0$ otherwise. 

The matrix $\mathcal{M}$ is almost a \emph{diagonal} matrix, except
for the non-zero elements in the row $j=i$; however, the
calculation of its eigenvalues is a simple matter. It can be proved
that the eigenvalues $\omega_l$ are solutions to the algebraic
equation
\begin{equation}
  \Pi^{n+1}_{l \neq i} \left[ 3\left( \gamma_{i} -
  \gamma_l \right) -\omega_l \right] = 0 \, , \label{eq:2}
\end{equation}
where $l$ runs through the permitted $(n+1)$ values, except for the
the particular value $l=i$, and then there are only $n$
eigenvalues.

In principle, the trivial critical point should be treated separately,
and the eigenvalue equation in this case is
\begin{equation}
  \Pi^{n}_{l=1} \left[ 3\left( \gamma_{n+1} - \gamma_l \right)
  -\omega_l \right] = 0 \, . \label{eq:4}
\end{equation}
Naively, Eq.~(\ref{eq:4}) seems to be a particular case of
Eq.~(\ref{eq:2}), one in which $i^\prime=n+1$. However, this is not
formally so, as the $n+1$-th perfect fluid does not appear in
Eqs.~(\ref{dynasysa}).

Therefore, we come to the following general conclusions.
\begin{itemize}
\item The critical point corresponding to the domination of the
  perfect fluid with the \emph{largest} equation of state is
  \emph{unstable}.

\item The critical point corresponding to the domination of the
  perfect fluid with the \emph{smallest} equation of state is
  \emph{stable}.

\item All other critical points are \emph{saddle} points.
\end{itemize}

The above statements directly follow from Eqs.~(\ref{eq:2})
and~(\ref{eq:4}), as the positivity or negativity of the perturbation
eigenvalues for a particular case depend on the relative value of the
dominant equation of state with respect to the others.

\section{How does the universe expand?}
\label{sec:expansion}
In this section, we will work on particular and simple models of the
universe, and draw their dynamics according to their material contents
and the methods discussed in the previous section.

\subsection{Our universe, or the Concordance Model}
\label{cm-universe}
We now turn our attention to the so-called \textit{concordance model}
(CM), also known as the $\Lambda$CDM (Lambda Cold Dark Matter) model,
the model of the universe the cosmological observations altogether
seem to favor\cite{Conley:2006qb,Lambda,wiki2}. It contains radiation
(relativistic particles), matter and vacuum energies in the
proportions $\Omega_{r,0}=10^{-5}$, $\Omega_{m,0}=0.266$,
$\Omega_{\Lambda ,0}=0.732$, and $\Omega_{k,0}= 0.01$, respectively. 

For numerical purposes, we take $\Omega_{r,0}=10^{-5}$,
$\Omega_{m,0}=0.3$, $\Omega_{\Lambda,0}=0.69$, and
$\Omega_{k,0}=0.01$, so that the Friedmann constraint is accurately
accomplished at the present time, see Eq.~(\ref{fried3}).

\subsubsection{Effective gravitational potential}
\begin{figure}
\includegraphics[width=8cm]{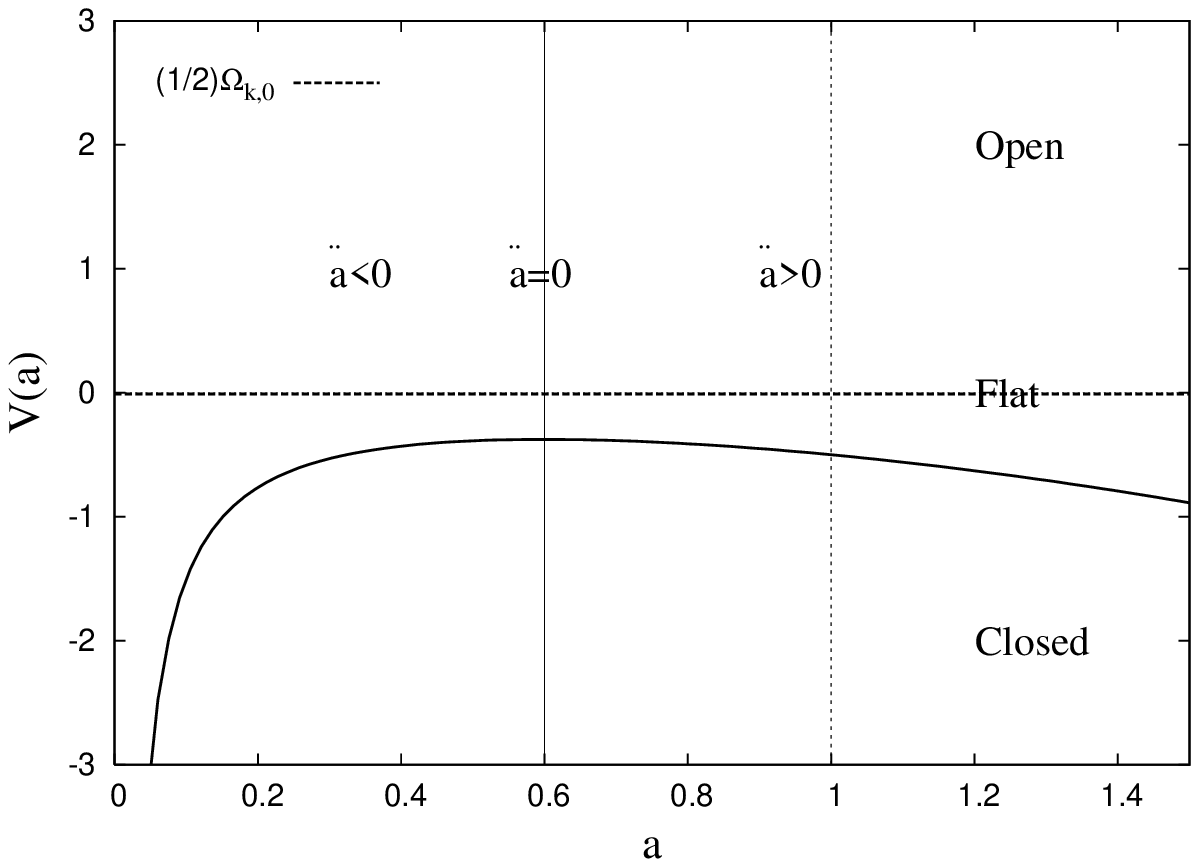}
\includegraphics[width=8cm]{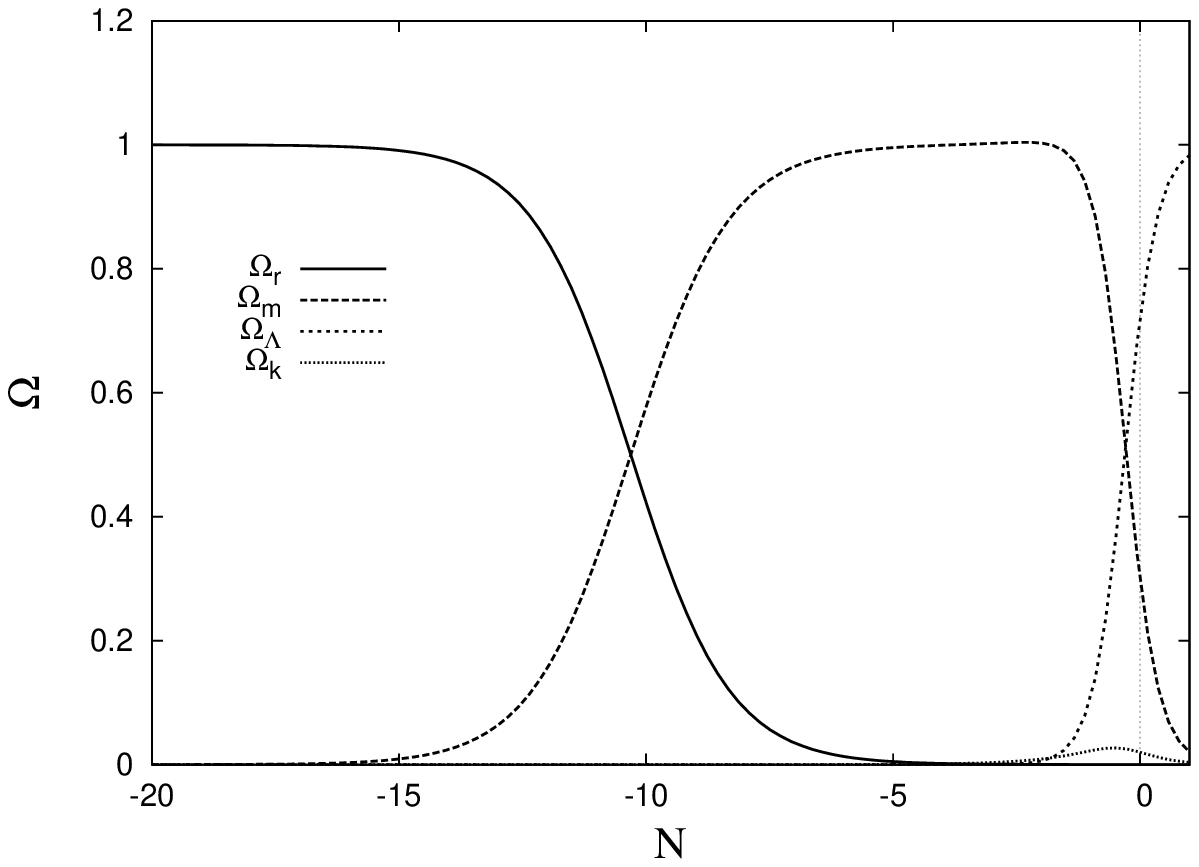}
\caption{\label{fig:gravpot-cm} (Top) The effective gravitational potential
  $V(a)$, see Eq.~(\ref{gravpot-cm}), for the so-called concordance
  model, according to the values of the density parameters $\Omega_i$
  obtained from recent cosmological observations. The solid vertical
  line goes through the maximum of the potential, and marks the
  time at which $\ddot{a}=0$. (Bottom) The evolution of the energy
  density parameters $\Omega_j$ as functions of the number of $e$-foldings $N$,
  see Eqs.~(\ref{dpsols}). Notice that there are radiation, matter and
  $\Lambda$ dominated stages; though the curvature is not zero, a
  curvature dominated era cannot be achieved. The dashed vertical line
  is at the present value of the (normalized) scale factor $a_0=1$ ($N=0$) in
  both figures.}
\end{figure}

The effective gravitational potential of the CM is
\begin{eqnarray}
V(a) &=& - \frac{1}{2} \left( \frac{\Omega_{r,0}}{a^2} +
\frac{\Omega_{m,0}}{a} + \Omega_{\Lambda ,0} a^2 \right) \, .
\label{gravpot-cm}
\end{eqnarray}
As the density parameters are positive definite, we see that the
gravitational potential has one critical point. As radiation does not
contribute significantly, the critical point is approximately at $a_c
\simeq \sqrt[3]{\Omega_{m,0}/(2\Omega_{\Lambda ,0})}=0.6$. As said before,
this critical point is a \textit{maximum}, as $V^{\prime \prime}(a_c)
= -\Omega_{r,0} a^{-4}_c -3 \Omega_{\Lambda ,0} < 0$. For
completeness, we plot $V(a)$ in Fig.~\ref{fig:gravpot-cm} for a CM
universe.

The curvature of the universe is so small that one hardly
distinguishes our universe from the 'Flat' case in
Fig.~\ref{fig:gravpot-cm}, and the maximum of the effective
gravitational potential is negative, $V(a_c) < 0$. These two facts
imply that the CM universe will never stop expanding, had a past
decelerating stage, and is currently accelerating its expansion. 

Notice that, keeping the radiation and matter present contributions
fixed, the larger the vacuum contribution is, the sooner the expansion
starts accelerating.

\subsubsection{Dynamical system}
Explicitly, the dynamical system of the CM is
\begin{subequations}
\label{cmds}
\begin{eqnarray}
\Omega^\prime_r &=& \Omega_r \left( 2\Omega_r + \Omega_m -2 \Omega_\Lambda -
2 \right) \, , \\
\Omega^\prime_m &=& \Omega_m \left( 2\Omega_r + \Omega_m -2 \Omega_\Lambda -
1 \right) \, , \\
\Omega^\prime_\Lambda &=& \Omega_\Lambda \left( 2\Omega_r + \Omega_m -2
\Omega_\Lambda +2 \right) \, ,
\end{eqnarray}
\end{subequations}
and its critical points, together with their respective stability
analysis, are shown in Table~\ref{t:critical}; the eigenvalues of the
perturbation matrix~(\ref{matrixpert}) are given in
Table~\ref{t:eigen}.

\begin{table}[htp]
\caption{ \label{t:critical} Critical points for the CM dynamical
  system~(\ref{cmds}). As the curvature is positive, the trivial
  solution does not exist for the CM model.}
\begin{ruledtabular}
\begin{tabular}{cccc}
Domination & $\Omega_k$ & $\mathbf{x}_0$ & Stability \\ \hline
Radiation & $0$ & $(1,0,0)$ & Unstable \\
Matter & $0$ & $(0,1,0)$ & Saddle \\
$\Lambda$ & $0$ & $(0,0,1)$ & Stable
\end{tabular}
\end{ruledtabular}
\end{table}

\begin{table}[htp]
\caption{ \label{t:eigen} Eigenvalues of the perturbation
  matrix~(\ref{matrixpert}) corresponding to the critical
  points shown in Table~\ref{t:critical}; see also Eqs.~(\ref{eq:3})
  and ~(\ref{eq:2}).}
\begin{ruledtabular}
\begin{tabular}{cccc}
Domination & $\omega_1$ & $\omega_2$ & $\omega_3$ \\ \hline
Radiation & $1$ & $2$ & $4$ \\
Matter & $-1$ & $1$ & $3$ \\
$\Lambda$ & $-2$ & $-3$ & $-4$
\end{tabular}
\end{ruledtabular}
\end{table}

As explained before, the trivial solution is also a solution of the
dynamical system~(\ref{cmds}), but its existence is forbidden on
physical grounds because of the Friedmann constraint.

The equations of state are $\gamma_r > \gamma_m > \gamma_\Lambda$, and
thus we conclude that radiation domination is an unstable point,
matter domination is a saddle point, and $\Lambda$ domination is the
only stable point. The universe will always reach a $\Lambda$ dominated era
at some epoch, and will remain in it thereafter. We want to point out
that this is true whatever the actual contributions of each component
are.

The evolution of the density parameters $\Omega_j$ is also shown in
Fig~\ref{fig:gravpot-cm}. As discussed before, one can see the
radiation, the matter and the $\Lambda$ dominated eras, but the
curvature dominated solution is never achieved, even though its
contribution is noticeable at the transition between the matter and
$\Lambda$ dominated eras.

\subsection{Einstein's static universe}
Originally, Einstein considered the universe as \textit{static}, and
he introduced a cosmological constant to allow his
equations~(\ref{einstein}) to have a static solution. That is,
Einstein found a solution with $\dot{a}=\ddot{a}=0$. For an
interesting discussion on Einstein's cosmological model see Chs.~2
and~4 in\cite{peebles93}.

\subsubsection{Effective gravitational potential}
According to Eqs.~(\ref{fried2}) and~(\ref{deriv1}), Einstein's static
solution corresponds to a universe located at exactly the critical
point $V^\prime (a_c)=0$, for which $V(a_c)=(1/2)\Omega_{k,0}$. That
is, the universe has just the enough total energy to be at the
critical point of its effective gravitational potential.

\begin{figure}
\includegraphics[width=8cm]{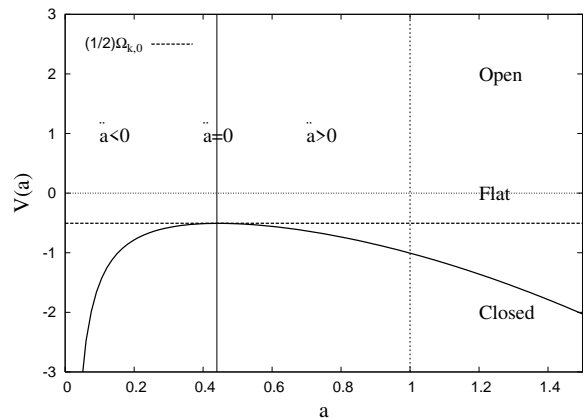}
\caption{\label{fig:einstein} The effective gravitational potential
  $V(a)$, see Eq.~(\ref{gravpot-cm}), for an Einstein's static
  universe with $\Omega_{m,0}=0.3$, $\Omega_{\Lambda ,0}=1.713$, and
  $\Omega_{k,0}=-1.013$; the critical point is at $a_c=0.44$. The above
  values are determined from the conditions $V(a_c)=(1/2)\Omega_{k,0}$,
  $V^\prime(a_c)=0$, and the Friedmann constraint~(\ref{fried3}).}
\end{figure}

Unfortunately, we have already learned that the critical point of
  $V(a)$ that Einstein found must be a maximum, and therefore is an
  \textit{unstable point}. Hence, the universe should expand or
  collapse, but cannot remain static! An example of a static universe
  with density parameters $\Omega_{m,0}=0.3$,
  $\Omega_{\Lambda ,0}=1.713$, $\Omega_{k,0}=-1.013$ is shown in
  Fig.~\ref{fig:einstein}. 

Some time after Einstein proposed its static universe, the expansion
of the universe was discovered, and Einstein thought he had made a
mistake in introducing a cosmological constant to have a static
universe; then called the latter his 'biggest blunder'. 

It took a bit longer for cosmologists to recognize that Einstein's
original model is unstable\cite{Eddington:1930,peebles93}, and that
the universe can expand even in the presence of a finely-tuned
cosmological constant.

\subsubsection{Dynamical system}
The dynamical system analysis proposed in a previous section can, in
principle, be applied to the Einstein's static universe. The
conclusion is that the critical points are the same as those of the
concordance model, see Table~\ref{t:critical} and~\ref{t:eigen}. This
is because the existence and nature of the critical points only depend
on the equations of state of the material content of the universe.

However, Einstein's universe is a case we have to deal with carefully,
because one cannot define a critical density at the maximum of the
potential $V(a_c)$, since the latter this time implies the vanishing
of the Hubble parameter $H$.

\subsection{Dark energy}
Since the discovery of the accelerated expansion of the universe in the
observations of Type Ia supernovae\cite{Conley:2006qb}, cosmologists
have been wondering whether the energy responsible for the
acceleration is a cosmological constant ($\gamma = 0$, $\rho_\Lambda =
\textrm{const}$), or another exotic kind of matter, usually dubbed
\textit{dark energy}, with an equation of state that lies in the range
$2/3 > \gamma_X > 0$ and $\rho_X \neq \textrm{const}$ (we shall use an
$X$ to denote dark energy
quantities)\cite{Conley:2006qb,Lambda,Padmanabhan:2006kz}.

For concreteness, we will next work on the case of a CM universe in
which vacuum energy is changed by a dark energy component with
a \emph{constant} equation of state $\gamma_X = 1/3$. It should be
warned that this is not the most general case of dark energy, but we
consider a constant $\gamma_X$ for pedagogical reasons only.

\subsubsection{Effective gravitational potential}
\begin{figure}[floatfix]
\includegraphics[width=8cm]{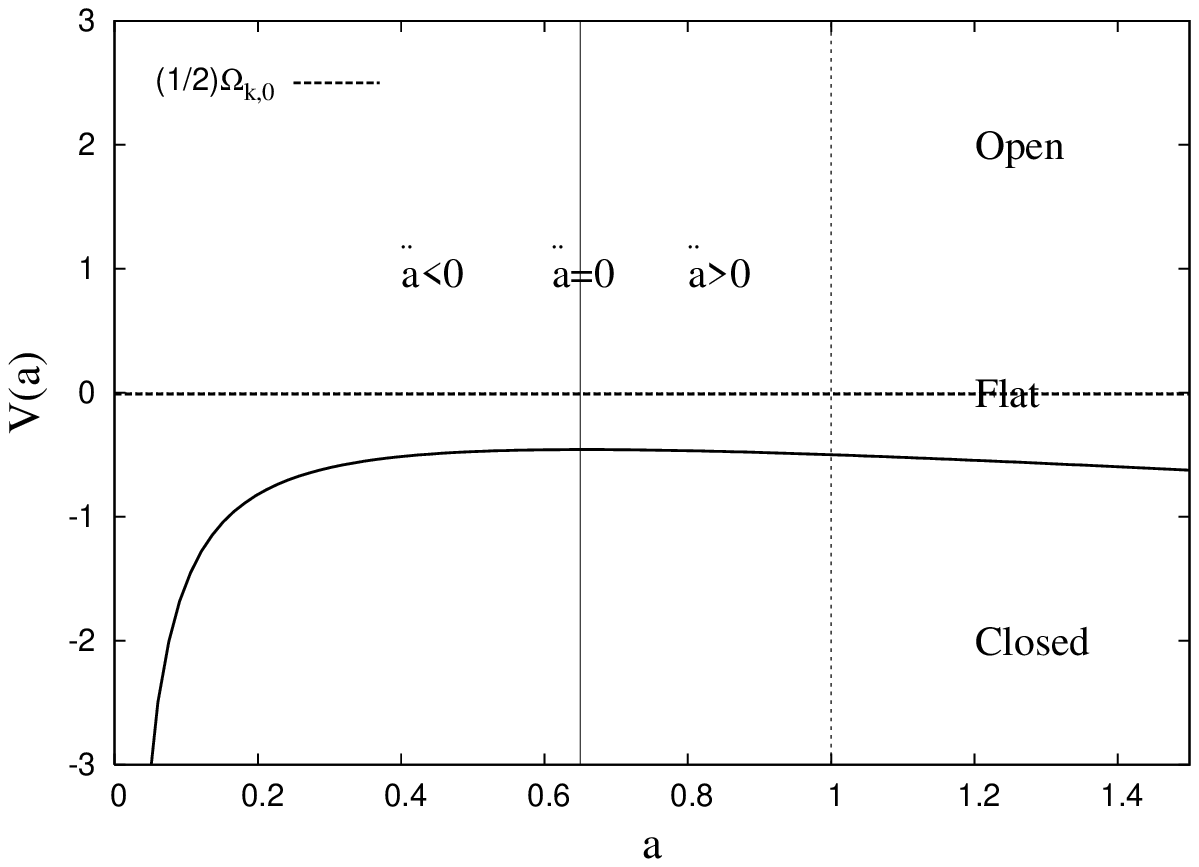}
\includegraphics[width=8cm]{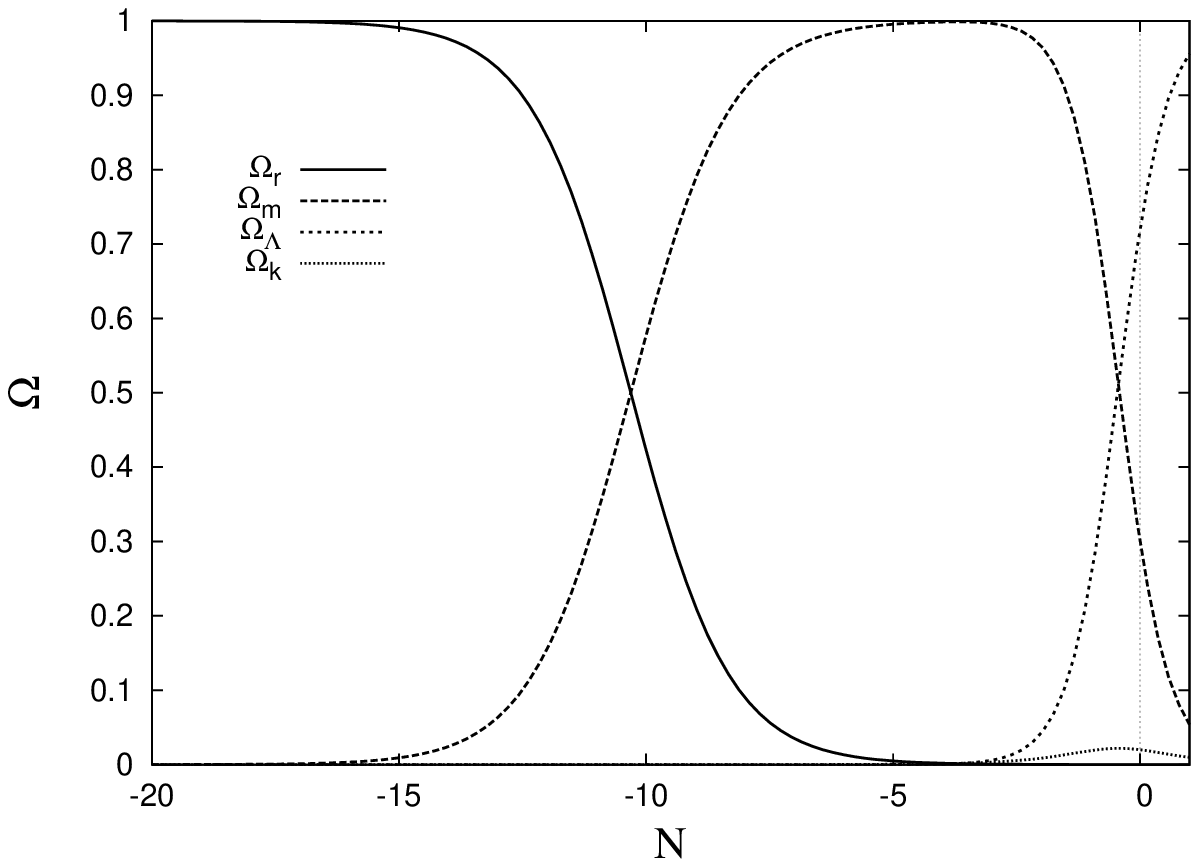}
\caption{\label{fig:de} (Top) The effective gravitational potential
  $V(a)$ for a CM universe in which the vacuum energy is replaced by a
  dark energy component with equation of state $\gamma_X =
  1/3$. (Bottom) The corresponding evolution of the density parameters
  $\Omega_j$, see Eqs.~(\ref{dpsols}), as functions of the number of
  $e$-foldings $N$. As in Fig.~\ref{fig:gravpot-cm}, there are
  radiation, matter and dark energy dominated stages, without a
  curvature one. The dashed vertical line
  is at the present value of the (normalized) scale factor $a_0=1$
  ($N=0$) in both figures.}
\end{figure}

The effective gravitational potential and the evolution of the density
parameters are shown in Fig.~\ref{fig:de}. As in the standard CM
model, there is a maximum in the effective gravitational potential,
but the behavior at late times differs from that of the CM. This time
the maximum of the potential is at $a_c \simeq
\sqrt{\Omega_{m,0}/\Omega_{X,0}}=0.65$. That is, it takes longer for
this universe to have a dark energy dominated, and then accelerated,
stage than in the case of a cosmological constant. This was to be
expected, as we mentioned before that a cosmological constant is an
extreme case for the equation of state of a perfect fluid.

\subsubsection{Dynamical system}
The dynamical system for a universe with a dark energy component is
\begin{subequations}
\label{deds}
\begin{eqnarray}
\Omega^\prime_r &=& \Omega_r \left[ 2\Omega_r + \Omega_m
- \Omega_X - 2 \right] \, , \\
\Omega^\prime_m &=& \Omega_m \left[ 2\Omega_r + \Omega_m
- \Omega_X - 1 \right] \, , \\
\Omega^\prime_X &=& \Omega_\Lambda \left[ 2\Omega_r + \Omega_m
- \left( \Omega_X - 1 \right) \right] \, ,
\end{eqnarray}
\end{subequations}
and the existence and nature of its critical points are shown in
Table~\ref{t:critical-de}.

There is only one stable solution, which is the dark energy dominated
one. This will make the universe have an accelerated expansion at late
times. We see that the model in this section is similar to the CM, but
an accurate enough measurement could in principle distinguish between
the two. So far, the CM seems to be the chosen one,
see\cite{Conley:2006qb,Lambda}.

\begin{table}[floatfix]
\caption{ \label{t:critical-de} Critical points for the dynamical
  system~(\ref{deds}) of a universe containing a dark energy
  component.}
\begin{ruledtabular}
\begin{tabular}{cccc}
Domination & $\Omega_k$ & $\mathbf{x}_0$ & Stability \\ \hline
Radiation & $0$ & $(1,0,0)$ & Unstable \\
Matter & $0$ & $(0,1,0)$ & Saddle \\
$X$ & $0$ & $(0,0,1)$ & Stable 
\end{tabular}
\end{ruledtabular}
\end{table}

\begin{table}[htp]
\caption{ \label{t:eigen-de} Eigenvalues of the perturbation
  matrix~(\ref{matrixpert}) corresponding to the critical
  points shown in Table~\ref{t:critical-de}.}
\begin{ruledtabular}
\begin{tabular}{cccc}
Domination & $\omega_1$ & $\omega_2$ & $\omega_3$ \\ \hline
Radiation & $1$ & $4$ & $3$ \\
Matter & $-1$ & $3$ & $2$ \\
$X$ & $-3$ & $-2$ & $-1$
\end{tabular}
\end{ruledtabular}
\end{table}

\section{Final comments}
\label{sec:conclusions}
We stated at the beginning of this manuscript our intention to
understand the solutions of Einstein's equations in the case of a
homogeneous and isotropic universe. After all of the mathematical work
done in the previous sections, we can extract some general conclusions
on how the expansion of the universe depend on its material content
and curvature.

Assuming all of the energy densities in the universe are
\textit{positive} definite, the effective gravitational
potential~(\ref{gravpot}) is \textit{negative} definite. Therefore,
\begin{enumerate}
\item An \textit{open} universe will expand for ever, irrespective of
  the material content and of how large is its negative curvature. If
  there is a component with an equation of state such that $\gamma <
  2/3$, an open universe will enter an accelerated stage at late
  times. Otherwise, it will decelerate but the expansion rate will
  never vanish.

\item A \textit{flat} universe will always expand, and will accelerate
  its expansion if there is a component with an equation of state such
  that $\gamma < 2/3$. Otherwise, the expansion rate will
  asymptotically vanish ($\dot{a} \rightarrow 0$ as $t \rightarrow
  \infty$).

\item A \textit{closed} universe will have an accelerated expansion if
  there is a component with an equation of state such that $\gamma <
  2/3$, and if its curvature is such that $\Omega_{k,0} <
  2V(a_c)$. If at least one of the previous conditions is not
  accomplished, then the universe will recollapse.
\end{enumerate}

Some other comments are in turn. It was shown how the dynamics of the
universe can be studied using two complementary approaches. For
instance, the gravitational potential method gave a clear proof of the
instability of the Einstein's static universe, whereas the dynamical
system method revealed the stability nature of the different
domination stages our universe has passed through. It is our hope that
the two methods revised in this paper can help undergraduate students
to deal with the expansion of a homogeneous and isotropic universe.

Moreover, we would like to stress the fact that such methods are taken
seriously for the analysis of a great variety of cosmological
models. Already mentioned was the use of the 'gravitational
potential' method in\cite{padh94,rindler01,Szydlowski:2006ma}. On the
other hand, the use of dynamical systems is widely known in the
specialized literature; for a non-exhaustive list of examples, see
Refs.\cite{dynasystems,Coley:1999uh,Copeland:1997et,Collinucci:2004iw,Urena-Lopez:2005zd,Copeland:2006wr}
and references there in.

In our opinion, it is important to teach undergraduate students
different methods and techniques currently used in specialized
research. For we cannot say to what extension the methods presented in
this work can be used and generalized to other equations in Cosmology;
but that we shall know from the work of future cosmologists.

\begin{acknowledgments}
I would like to thank two anonymous referees for their useful
suggestions and comments that helped to improve this manuscript. This
work was partially supported by grants from CONACYT (42748, 46195,
47641), CONCYTEG 05-16-K117-032, DINPO 85, and PROMEP UGTO-CA-3.
\end{acknowledgments}

\bibliography{expansionrefs}

\end{document}